\documentclass[12pt,a4paper]{amsart}

\usepackage[headings]{fullpage}
\usepackage{amsmath,amssymb}
\usepackage{microtype}
\usepackage{needspace}
\UseMicrotypeSet[expansion]{basictext}
\usepackage[hidelinks]{hyperref}

\hypersetup{
  pdftitle={Maximum purity of absolutely separable and absolutely PPT states
    in qutrit--qudit systems},
  pdfauthor={Anh T. Tran},
  pdfsubject={Maximum purity of absolutely separable and APPT states in
    qutrit--qudit systems},
  pdfkeywords={absolutely separable states, absolutely PPT states,
    qutrit--qudit systems, purity, spectral optimization, depolarizing noise}
}

\numberwithin{equation}{section}

\theoremstyle{plain}
\newtheorem{theorem}{Theorem}[section]
\newtheorem{lemma}[theorem]{Lemma}
\newtheorem{proposition}[theorem]{Proposition}
\newtheorem{corollary}[theorem]{Corollary}
\newtheorem{maintheorem}{Theorem}

\newcommand{\BC}{\mathbb{C}}
\newcommand{\CA}{\mathcal{A}}
\newcommand{\CT}{\mathcal{T}}
\newcommand{\CO}{\mathcal{O}}

\begin{document}

\title[AS and APPT purity in qutrit--qudit systems]%
  {Maximum purity of absolutely separable and absolutely PPT states in
  qutrit--qudit systems}

\author{Anh T. Tran}
\thanks{Department of Mathematical Sciences, The University of Texas at Dallas,
  Richardson, TX 75080, USA\@. Corresponding author:
  \href{mailto:att140830@utdallas.edu}{att140830@utdallas.edu}.}


\begin{abstract}
  A bipartite state is absolutely positive partial transpose (APPT) if it
  remains PPT under every global unitary transformation, and absolutely
  separable if it remains separable.  For every qutrit--qudit system
  $\BC^3\otimes\BC^n$, $n\ge3$, we prove that the maximum purity is the same
  for absolutely separable and APPT states and equals
  $\max\{(3n+8)/(3n+2)^2,\,3/(8n)\}$.  The maximizing ordered spectrum is
  unique and contains exactly two distinct eigenvalues: the larger one has
  multiplicity one for $3\le n\le8$ and multiplicity $n$ for $n\ge9$.
  We also derive, in terms of the purity, an upper bound on the minimum
  partial-transpose eigenvalue achievable by global unitary transformations,
  and characterize all states for which the bound is attained.  These results
  confirm the inscribed-polytope purity conjecture of Ahiable, Kothakonda, and
  Winter for all qutrit--qudit systems.  They also give a lower bound, sharp
  over explicit purity ranges, on the global depolarizing noise needed for
  absolute separability or APPT.
\end{abstract}
\maketitle

\section{Introduction}

A bipartite quantum state on $\BC^m\otimes\BC^n$ is a positive
semidefinite operator $\rho$ of trace one.  It is \emph{separable} if it can
be written as a convex combination of product states,
\[
  \rho=\sum_j p_j\,\rho_j^A\otimes\rho_j^B,
  \qquad p_j\ge0,\qquad \sum_jp_j=1,
\]
where $\rho_j^A$ and $\rho_j^B$ are density operators on the two subsystems
\cite{Pe}.  Such a state can be prepared locally using shared classical
randomness.  A state that admits no such decomposition is \emph{entangled}.

The positive partial transpose (PPT) criterion provides a necessary test
for separability.  Writing
$\rho^\Gamma=(\operatorname{id}\otimes T)(\rho)$, where $T$ denotes matrix
transposition on the second subsystem, a state is PPT if
$\rho^\Gamma\succeq0$.  Every separable state is PPT \cite{Pe}, so a
negative eigenvalue of $\rho^\Gamma$ certifies entanglement.  The converse
holds for $2\otimes2$ and $2\otimes3$, but fails in general~\cite{HHH}.

Local unitary transformations preserve separability, whereas global
unitaries can create entanglement without changing the spectrum.  This
leads to a stronger question: which states remain separable under every
global unitary transformation?  Such states are called \emph{absolutely
  separable}.  The corresponding PPT requirement defines \emph{absolutely
  positive partial transpose} (APPT) states: $\rho$ is APPT if
$(U\rho U^\dagger)^\Gamma\succeq0$ for every global unitary $U$ \cite{Jo}.
Every absolutely separable state is therefore APPT.  Both absolute
properties depend only on the spectrum, unlike separability and PPT for
an individual state.  Hildebrand characterized APPT spectra by finitely
many matrix inequalities \cite{Hi}.

For an ordered spectrum $\boldsymbol\lambda=(\lambda_1,\ldots,\lambda_d)$,
where $d=mn$, the purity is
\[
  F(\boldsymbol\lambda)=\operatorname{Tr}(\rho^2)
  =\sum_{i=1}^d\lambda_i^2.
\]
Purity quantifies mixedness and, by itself, does not determine whether a
state is entangled.  We first seek the largest purity compatible with being
PPT under every global unitary and identify all spectra attaining it.  We then
show that the same maximum and the same maximizing spectra hold for absolute
separability.
Above this common maximum, some global unitary must produce a state with
negative partial transpose, and hence an entangled state.  Purity below the
maximum, however, does not guarantee either absolute property.

For qubit--qudit systems, APPT and absolute separability coincide
\cite{Jo}, and the maximum purity is already known.  Z.~Song and L.~Chen
characterized the extreme points and obtained the maximum purity
\cite[Theorems~9--10 and Corollary~11(ii),(iv)]{SC}: it is $3/8$ for
$2\otimes2$, while for $2\otimes n$ with $n\ge3$ it equals
\[
  \begin{cases}
    \dfrac{2}{3n}, & n\text{ even},\\[6pt]
    \dfrac{6n+4}{(3n+1)^2}, & n\text{ odd}.
  \end{cases}
\]
Maximizing purity over the outer polytope in \cite[Theorem~A]{TrS}
independently recovers these values for $n\ge4$.

For general systems $\BC^m\otimes\BC^n$ with $2\le m\le n$ and $mn\ge6$,
\cite[Theorem~A]{TrS} gives an explicit upper bound depending only on
$mn$, asymptotic to $4/(3mn)$.  It is attained in qubit--qudit systems,
but is strict in every qutrit--qudit system with $n\ge3$: the unique
maximizing ordered spectrum of the outer polytope is not APPT
\cite[Theorem~A and Remark~1.1]{TrS}.  Thus the maximum over this larger
polytope does not determine the qutrit APPT maximum.

Z.~Song and L.~Chen reported numerical evidence that the maximum purity of
two-qutrit APPT states is $17/121$, attained at the spectrum
$\frac1{11}(3,1,\ldots,1)$ \cite[discussion following Corollary~16]{SC}.
The same value was later stated as an exact maximum by M.~Chen and N.~Klco
\cite[Theorem~II.6]{CK}.  Both proofs of that theorem in version~1
rely on the incorrect lower bound in Lemma~II.4.\footnote{%
  \interfootnotelinepenalty=10000
  Lemma~II.4 of \cite{CK} asserts that the largest eigenvalue $\lambda_1$
  of a two-qutrit density matrix with purity $P$ satisfies
  $\lambda_1\ge(1+\sqrt{8(9P-1)})/9$.  For a counterexample, consider
  \[
    \rho_0=\frac1{12}\operatorname{diag}(2,2,2,1,1,1,1,1,1).
  \]
  It has purity $P=\operatorname{Tr}(\rho_0^2)=1/8$ and largest eigenvalue
  $\lambda_1(\rho_0)=1/6$, contradicting the asserted inequality.}
Consequently, those arguments do not establish the asserted purity bound.
We prove the same two-qutrit value independently as part of a complete
qutrit--qudit theorem covering every $n\ge3$ and determine all maximizing
spectra.  Relatedly, N.~Wang, L.~Chen, and Z.~Song recently analyzed the
boundary and extreme points of full-rank two-qutrit APPT states with exactly
three distinct eigenvalues \cite{WCS}.

Ahiable, Kothakonda, and Winter determined the maximum purity on a
permutation-invariant polytope of APPT spectra
\cite[Theorems~6.2--6.3]{AKW}.  Its maximum gives a lower bound on the
maximum APPT purity, but does not give a matching upper bound.
They conjectured that the two maxima agree \cite[Conjecture~6.7]{AKW}.
We establish this equality for every qutrit--qudit system and characterize
all maximizing spectra.  Moreover, although a complete spectral
characterization of absolute separability is unknown in these dimensions,
known sufficient criteria are enough to certify both maximizing spectra as
absolutely separable by \cite[Theorem~3]{AV} and \cite[Theorem~1]{Kondra}.
Hence the same maximum and equality cases hold for absolute separability.

\begin{samepage}
From now on, $n\ge3$ and $d=3n$.  For multiplicities, $x^{[q]}$ denotes
$q$ consecutive entries equal to $x$, with no entries when $q=0$.
For these dimensions, the two candidate spectra from the
Ahiable--Kothakonda--Winter polytope are
\[
  \boldsymbol\lambda^{\mathrm{one}}
  =\frac1{3n+2}(3,1^{[3n-1]}),
  \qquad
  \boldsymbol\lambda^{\mathrm{cen}}
  =\frac1{4n}(2^{[n]},1^{[2n]}),
\]
\end{samepage}
with purities
\[
  M_{\mathrm{one}}:=\frac{d+8}{(d+2)^2},\qquad
  M_{\mathrm{cen}}:=\frac9{8d},\qquad
  M:=\max\{M_{\mathrm{one}},M_{\mathrm{cen}}\}.
\]
Their difference is
\begin{equation}
  \label{eq:qutrit-two-candidate-gap}
  M_{\mathrm{one}}-M_{\mathrm{cen}}
  =\frac{160-(d-14)^2}{8d(d+2)^2}.
\end{equation}
Since $d=3n$, the first candidate has larger purity for $3\le n\le8$,
and the second for $n\ge9$.  The following theorem shows that these
candidate values are sharp.

\begin{maintheorem}
  \label{thm:qutrit}
  For every integer $n\ge3$, the maximum purity of absolutely separable
  states and of APPT states on $\BC^3\otimes\BC^n$ is
  \begin{equation}
    \label{eq:qutrit-maximum}
    \begin{cases}
      \dfrac{3n+8}{(3n+2)^2}, & 3\le n\le8,\\[6pt]
      \dfrac{3}{8n},          & n\ge9.
    \end{cases}
  \end{equation}
  The unique maximizing ordered spectra for $3\le n\le8$ and $n\ge9$ are,
  respectively,
  \begin{equation}
    \label{eq:qutrit-maximizers}
    \frac{1}{3n+2}(3,1^{[3n-1]})
    \quad\text{and}\quad
    \frac{1}{4n}(2^{[n]},1^{[2n]}).
  \end{equation}
  In either class, the maximizing states are precisely the unitary conjugates
  of a diagonal state with the corresponding displayed spectrum.
\end{maintheorem}

\begin{corollary}
  \label{cor:qutrit-AKW}
  Conjecture~6.7 of Ahiable, Kothakonda, and Winter \cite{AKW} holds for
  $m=3$ and every $n\ge3$.  In particular, the maximum purity of
  $\mathrm{APPT}_{3,n}$ equals the maximum purity of their inscribed polytope
  $\mathcal P_{3,n}$, and the maximizing ordered spectra coincide.
\end{corollary}

\begin{proof}
  The maximum purity and the maximizing spectra of $\mathcal P_{3,n}$ are given
  in \cite[Theorem~6.3 and Corollary~6.4]{AKW}.  Theorem~\ref{thm:qutrit}
  shows that, in every qutrit--qudit dimension, the corresponding spectrum is
  also the unique maximizing ordered APPT spectrum, with the same purity.
\end{proof}

\begin{samepage}
\smallskip \noindent\textbf{Quantitative consequences.}
The purity maximum yields two further bounds.  Let $\rho$ have purity
$P=\operatorname{Tr}(\rho^2)$, and let $M$ be the maximum purity in
Theorem~\ref{thm:qutrit}.  First, the smallest partial-transpose
eigenvalue attainable by a global unitary satisfies
\[
  \min_U\lambda_{\min}\!\left((U\rho U^\dagger)^\Gamma\right)
  \le\frac1d\left(1-\sqrt{\frac{P-1/d}{M-1/d}}\right),
\]
where $U$ ranges over all unitary operators on $\BC^d$.
For $3\le n\le8$, the eigenvalue bound is attained at every purity
$P\in[1/d,1]$.  For $n\ge9$, it is attained precisely at purities
$P\in[1/d,1/n]$.

Second, consider the depolarized state $(1-p)\rho+(p/d)I_d$, where
$0\le p\le1$.  If $P>M$ and this state is absolutely separable or APPT,
then
\[
  p\ge1-\sqrt{\frac{M-1/d}{P-1/d}}.
\]
For either absolute property, the noise bound is attained at every purity
$M<P\le1$ when $3\le n\le8$.  For $n\ge9$, it is attained precisely at
purities $M<P\le1/n$.
Corollaries~\ref{cor:qutrit-min-pt}
and~\ref{cor:qutrit-depolarizing-purity} give the complete statements
and characterize all equality cases.
\end{samepage}

\smallskip \noindent\textbf{Proof strategy.}
The inequality $\lambda_1\le\lambda_{d-2}+\lambda_{d-1}+\lambda_d$
defines the outer spectral polytope studied in \cite{TrS}.  We first impose
the two elementary Hildebrand tests $G\ge0$ and $H\ge0$.  We compress the
$H=0$ face to a four-variable optimization.  The resulting auxiliary
polytope has one additional extremal direction.
At the corresponding spectrum, the Hildebrand matrix has a negative
eigenvalue with eigenvector $(\sqrt5,2,1)^\top$.  Testing with this vector
supplies a third inequality that removes the obstruction and explains the
appearance of the coefficient $\sqrt5$.
Bounds at the remaining vertices then give the APPT maximum and all equality
cases for every $n\ge3$.
The absolute-separability statement follows by certifying the two maximizing
spectra with known spectral criteria.

\smallskip \noindent\textbf{Organization.}
Section~\ref{sec:preliminaries} fixes the notation and proves the preliminary
geometric and edge bounds.  Section~\ref{sec:qutrit} contains the
four-variable purity bound, the additional Hildebrand inequality, and the proof
of Theorem~\ref{thm:qutrit}.  Section~\ref{sec:quantitative-consequence}
gives the geometric interpretation and proves the purity-dependent
partial-transpose eigenvalue and depolarizing-noise bounds, including all
equality cases.
Section~\ref{sec:conclusion}
summarizes the structural transition at $n=9$ and the remaining higher
local-dimension problem.

\section{Spectral notation and preliminary bounds}
\label{sec:preliminaries}

\subsection{Ordered spectra and face dimensions}
\label{sec:preliminaries-geometry}

Let
\[
  \CO_d:=\left\{\boldsymbol\lambda\in\mathbb R^d:
  \lambda_1\ge\cdots\ge\lambda_d\ge0,\quad
  \sum_{i=1}^d\lambda_i=1\right\},
\]
and write $\CA_{3,n}$ for its subset of APPT spectra on
$\BC^3\otimes\BC^n$.  Set $\lambda_{d+1}=0$ and define the coordinates
$y_i=\lambda_i-\lambda_{i+1}$.  Telescoping gives
\[
  \lambda_j=\sum_{i=j}^d y_i,\qquad
  \sum_{i=1}^d i y_i=1.
\]
Conversely, these formulas recover an ordered spectrum from any nonnegative
values $y_i$ with $\sum i y_i=1$.  Thus the map
$\boldsymbol\lambda\mapsto(y_1,2y_2,\ldots,d y_d)$ identifies $\CO_d$ with
the probability simplex.  Its vertices are therefore
\[
  \boldsymbol\nu_i
  =\bigl((1/i)^{[i]},0^{[d-i]}\bigr),\qquad 1\le i\le d,
\]
and every ordered spectrum has the unique representation
\begin{equation}
  \label{eq:gap-representation}
  \boldsymbol\lambda=\sum_{i=1}^d i y_i\boldsymbol\nu_i,
  \qquad y_i\ge0,\qquad \sum_{i=1}^d i y_i=1.
\end{equation}
The indices with $y_i>0$ identify the vertices of the smallest simplex face
containing $\boldsymbol\lambda$.  Square brackets denote convex hulls.
The rank of a spectrum is the number of its positive entries.

For distinct spectra $\boldsymbol\lambda,\boldsymbol\mu$ and $0<t<1$,
\[
  F((1-t)\boldsymbol\lambda+t\boldsymbol\mu)
  =(1-t)F(\boldsymbol\lambda)+tF(\boldsymbol\mu)
  -t(1-t)\|\boldsymbol\lambda-\boldsymbol\mu\|_2^2.
\]
Hence purity is strictly convex: every maximizer on a polytope is a vertex,
and every interior point of a nontrivial segment has purity strictly below
the larger endpoint purity.

An inequality is \emph{active} at a point if equality holds there.

\begin{samepage}
\begin{lemma}
  \label{lem:active-face-dimension}
  Let $R_0$ be a polytope and
  $P=R_0\cap\{\ell_1\ge0,\ldots,\ell_s\ge0\}$, where the $\ell_j$ are
  linear functionals.  For $\boldsymbol\lambda\in P$, let $Q_0$ be the unique
  face of $R_0$ whose relative interior contains $\boldsymbol\lambda$, and let
  $Q_P$ be the unique face of $P$ whose relative interior contains
  $\boldsymbol\lambda$.
  If the functionals corresponding to active added inequalities span a space
  of dimension $r$, then
  \begin{equation}
    \label{eq:face-dimension-bound}
    \dim Q_0\le\dim Q_P+r.
  \end{equation}
  If $R_0=\CO_d$ and $q=\dim Q_P$, then $\boldsymbol\lambda$ has at most
  $q+r+1$ distinct positive eigenvalues.  Intersecting with a single halfspace
  creates new vertices only in the relative interiors of old edges.
\end{lemma}
\end{samepage}

\begin{proof}
  Let $L$ be the direction space of $Q_0$, and let
  \[
    V:=\{w\in L:\ell_j(w)=0\text{ for every active }\ell_j\}.
  \]
  The active functionals impose at most $r$ independent conditions on $L$,
  so
  \[
    \dim V\ge\dim Q_0-r.
  \]

  Fix $w\in V$.  Since $\boldsymbol\lambda$ is in the relative interior of
  $Q_0$, for sufficiently small $\varepsilon>0$, both
  $\boldsymbol\lambda\pm\varepsilon w$ remain in $Q_0$.  The active added
  inequalities remain equalities, while the inactive ones remain strict;
  hence both perturbed points belong to $P$.  Since
  $\boldsymbol\lambda$ is their midpoint and belongs to the face $Q_P$,
  the defining property of a face forces both endpoints to lie in $Q_P$.
  Thus $V$ is contained in the direction space of $Q_P$, and therefore
  \[
    \dim Q_0-r\le\dim V\le\dim Q_P,
  \]
  proving \eqref{eq:face-dimension-bound}.

  For $R_0=\CO_d$, \eqref{eq:gap-representation} gives exactly
  $\dim Q_0+1$ indices with $y_i>0$, corresponding to the distinct positive
  eigenvalues in $\boldsymbol\lambda$.  The preceding bound proves
  the second assertion.

  Finally, a new vertex after one additional inequality has
  $\dim Q_P=0$, $r\le1$, and $\dim Q_0\ge1$.  Thus $\dim Q_0=1$,
  so it lies in the relative interior of an old edge.
\end{proof}

\subsection{The qutrit criterion and two inequalities}
\label{sec:qutrit-criterion}

The scalar inequalities used in the upper-bound proof all come from the matrix
\begin{equation}
  \label{eq:qutrit-K}
  K(\boldsymbol\lambda):=
  \begin{pmatrix}
    2\lambda_d & \lambda_{d-1}-\lambda_1 & \lambda_{d-3}-\lambda_2\\
    \lambda_{d-1}-\lambda_1 & 2\lambda_{d-2} & \lambda_{d-4}-\lambda_3\\
    \lambda_{d-3}-\lambda_2 & \lambda_{d-4}-\lambda_3 & 2\lambda_{d-5}
  \end{pmatrix}.
\end{equation}
Let $\widetilde K(\boldsymbol\lambda)$ be obtained by interchanging
$\lambda_{d-2}$ and $\lambda_{d-3}$ in \eqref{eq:qutrit-K}.
Hildebrand's qutrit criterion \cite{Hi} (see also \cite{SC})
gives, for $\boldsymbol\lambda\in\CO_d$,
\[
  \boldsymbol\lambda\in\CA_{3,n}
  \quad\Longleftrightarrow\quad
  K(\boldsymbol\lambda)\succeq0\quad\text{and}\quad
  \widetilde K(\boldsymbol\lambda)\succeq0.
\]
We use only $K\succeq0$ for the upper bound.
For fixed $\boldsymbol z\in\mathbb R^3$, the functional
\begin{equation}
  \label{eq:qutrit-scalar-functional}
  \ell_{\boldsymbol z}(\boldsymbol\lambda)
  :=\frac12\boldsymbol z^\top K(\boldsymbol\lambda)\boldsymbol z
\end{equation}
is linear in the spectrum and nonnegative on $\CA_{3,n}$.  Testing with
$(1,1,0)^\top$ and $(1,1,1)^\top$ gives, respectively,
\begin{equation}
  \label{eq:qutrit-initial-tests}
  \begin{aligned}
    G(\boldsymbol\lambda)
    &:=\lambda_{d-2}+\lambda_{d-1}+\lambda_d-\lambda_1,\\
    H(\boldsymbol\lambda)
    &:=\sum_{j=0}^{5}\lambda_{d-j}-(\lambda_1+\lambda_2+\lambda_3).
  \end{aligned}
\end{equation}
Thus $G,H\ge0$ on $\CA_{3,n}$, giving the initial outer polytope
\[
  R:=\{\boldsymbol\lambda\in\CO_d:G\ge0,\ H\ge0\},
  \qquad \CA_{3,n}\subseteq R.
\]
We omit the argument $\boldsymbol\lambda$ in linear inequalities when
no ambiguity can arise.

\smallskip \noindent\textbf{The candidate spectra.}
Since $n\ge3$, both candidates in
\eqref{eq:qutrit-maximizers} have their last six entries equal.
Thus $\lambda_{d-2}=\lambda_{d-3}$, and $K=\widetilde K$ at each
candidate.  Direct substitution gives
\[
  (d+2)K(\boldsymbol\lambda^{\mathrm{one}})
  =\begin{pmatrix}
    2&-2&0\\
    -2&2&0\\
    0&0&2
  \end{pmatrix},\qquad
  4nK(\boldsymbol\lambda^{\mathrm{cen}})
  =\begin{pmatrix}
    2&-1&-1\\
    -1&2&-1\\
    -1&-1&2
  \end{pmatrix}.
\]
Their quadratic forms are, respectively,
\[
  2(x_1-x_2)^2+2x_3^2,
  \qquad
  (x_1-x_2)^2+(x_1-x_3)^2+(x_2-x_3)^2.
\]
Both are nonnegative for every $x\in\mathbb R^3$, so the full qutrit
criterion proves that both spectra are APPT.
In particular, the maximum APPT purity is at least $M$.
Moreover,
\[
  (G,H)(\boldsymbol\lambda^{\mathrm{one}})
  =\frac1{d+2}(0,1),\qquad
  (G,H)(\boldsymbol\lambda^{\mathrm{cen}})
  =\frac1{4n}(1,0).
\]
The first candidate lies on $G=0$, while the central candidate lies on
$H=0$.  It remains to prove the matching upper bound and determine the
equality cases.

\subsection{Simplex and edge bounds}
\label{sec:qutrit-prelim}

Counting the nonzero entries at $\boldsymbol\nu_i$ gives
\begin{equation}
  \label{eq:qutrit-simplex-values}
  \begin{aligned}
    G(\boldsymbol\nu_i)
    &=\frac{\max\{0,i-d+3\}-1}{i},\\
    H(\boldsymbol\nu_i)
    &=\frac{\max\{0,i-d+6\}-\min\{i,3\}}{i}.
  \end{aligned}
\end{equation}
These are nonnegative exactly when $i\ge d-2$ and $i\ge d-3$,
respectively.  Therefore the simplex vertices in $R$ are
\[
  \boldsymbol\nu_{d-2},\quad\boldsymbol\nu_{d-1},\quad\boldsymbol\nu_d,
\]
whose largest purity is $1/(d-2)$.
If $\boldsymbol\lambda\in R$ has $\lambda_{d-1}=0$, then
$G\ge0$ gives $\lambda_1\le\lambda_{d-2}$.  Ordering and normalization
force $\boldsymbol\lambda=\boldsymbol\nu_{d-2}$.  Thus every other spectrum
in $R$ has rank $d-1$ or $d$.

The $G=0$ edge families are classified in \cite[Lemma~2.1]{TrS}.
We include their derivation and optimize with the additional constraint
$H\ge0$.

\begin{proposition}
  \label{prop:qutrit-G-edges}
  Every $\boldsymbol\lambda$ in the relative interior of an edge of $\CO_d$
  with $G=0$ and $H\ge0$ satisfies
  \begin{equation}
    \label{eq:qutrit-G-edge-bound}
    F(\boldsymbol\lambda)\le M_{\mathrm{one}}=\frac{d+8}{(d+2)^2}.
  \end{equation}
  Equality holds only at
  $\boldsymbol\lambda^{\mathrm{one}}=(3,1^{[d-1]})/(d+2)$.
\end{proposition}

\begin{proof}
  Suppose $[\boldsymbol\nu_i,\boldsymbol\nu_j]$, $i<j$, meets $G=0$
  in its relative interior.  By \eqref{eq:qutrit-simplex-values},
  $1\le i\le d-3$ and $j\in\{d-1,d\}$.  Write the intersection as
  $(a^{[i]},b^{[j-i]},0^{[d-j]})$ with $a>b>0$.
  The three smallest entries are $(b,b,0)$ or $(b,b,b)$, so $G=0$
  gives $a=2b$ or $a=3b$, respectively.  Normalization yields
  \[
    \frac{(2^{[i]},1^{[d-1-i]},0)}{d-1+i}\quad(j=d-1),
    \qquad
    \frac{(3^{[i]},1^{[d-i]})}{d+2i}\quad(j=d).
  \]
  To impose $H\ge0$, put
  $h_i=-iH(\boldsymbol\nu_i)=\min\{i,3,d-3-i\}$.
  The representation
  $\boldsymbol\lambda=i(a-b)\boldsymbol\nu_i+jb\boldsymbol\nu_j$ gives
  \[
    H(\boldsymbol\lambda)=(j-d+3)b-h_i(a-b).
  \]
  Thus $h_i\le2$ for $j=d-1$ and $h_i\le3/2$ for $j=d$.  Equivalently,
  the indices satisfying $H\ge0$ are
  \[
    i\le2\ \text{or}\ i\ge d-5\quad(j=d-1),
    \qquad
    i=1\ \text{or}\ i\ge d-4\quad(j=d).
  \]

  The purities of these two families are, respectively,
  \[
    \Phi_{d-1}(i)=\frac{d-1+3i}{(d-1+i)^2},\qquad
    \Phi_d(i)=\frac{d+8i}{(d+2i)^2}.
  \]
  For the full-rank family, direct subtraction gives
  \[
    M_{\mathrm{one}}-\Phi_d(i)
    =\frac{4(i-1)\bigl((d+8)i-d(d-1)\bigr)}
    {(d+2)^2(d+2i)^2}.
  \]
  This vanishes at $i=1$.  For the remaining indices $i\ge d-4$,
  the second factor in the numerator is at least $5d-32>0$.
  Thus $\Phi_d(i)\le M_{\mathrm{one}}$, with equality only at $i=1$.

  For the family of rank $d-1$, compare with the value at $i=2$:
  \[
    \frac{d+5}{(d+1)^2}-\Phi_{d-1}(i)
    =\frac{(i-2)\bigl((d+5)i-(d-1)(d-3)\bigr)}
    {(d+1)^2(d-1+i)^2}.
  \]
  For $i\le2$, the second factor is at most $-(d-7)(d+1)<0$;
  for $i\ge d-5$, it is at least $4(d-7)>0$.
  The product is therefore nonnegative on both stated index ranges,
  with equality only at $i=2$.  Hence
  \[
    \Phi_{d-1}(i)\le\frac{d+5}{(d+1)^2}<M_{\mathrm{one}},
  \]
  where the strict comparison follows from
  \[
    M_{\mathrm{one}}-\frac{d+5}{(d+1)^2}
    =\frac{(d-9)(d+2)+6}{(d+1)^2(d+2)^2}>0
    \qquad(d\ge9).
  \]
  Thus $\boldsymbol\lambda^{\mathrm{one}}$ is the unique maximizing spectrum.
\end{proof}

\section{Proof of Theorem A}
\label{sec:qutrit}

\subsection{The \texorpdfstring{$H=0$}{H=0} face}

By \eqref{eq:face-dimension-bound}, every vertex of $R$ with $H>0$ is a
simplex vertex or a $G=0$ edge point.  For $n\ge4$, all three simplex
vertices from Subsection~\ref{sec:qutrit-prelim} have purity strictly below
$M_{\mathrm{one}}$, since $d\ge12$ and
\[
  M_{\mathrm{one}}-\frac1{d-2}=\frac{2(d-10)}{(d-2)(d+2)^2}>0.
\]
When $n=3$, only $\boldsymbol\nu_7$ has purity above
$M_{\mathrm{one}}=17/121$; the other two simplex vertices have purity at most
$1/8<M_{\mathrm{one}}$.  The bound \eqref{eq:qutrit-G-edge-bound} applies
to the $G=0$ edge points, with equality only at
$\boldsymbol\lambda^{\mathrm{one}}$.

Thus the only vertices of $R$ not yet covered by these bounds are those on
$H=0$ and, when $n=3$, $\boldsymbol\nu_7$.

\begin{samepage}
\begin{lemma}
  \label{lem:qutrit-H-compression}
  Let $\boldsymbol\lambda\in R$ satisfy $H=0$.  Set
  \[
    h:=\lambda_1+\lambda_2+\lambda_3,
    \qquad
    v:=\frac{\lambda_{d-5}+\lambda_{d-4}+\lambda_{d-3}}3,
  \]
  and $\tau:=2h+(d-9)v$.  Then
  \begin{equation}
    \label{eq:qutrit-basic-compression-ineqs}
    \lambda_1\ge\lambda_2\ge\lambda_3\ge v>0,
    \qquad h\le6v,
    \qquad 3v\le\lambda_2+\lambda_3,
    \qquad \tau\le1,
  \end{equation}
  and
  \begin{equation}
    \label{eq:qutrit-compression-bound}
    F\le\sum_{i=1}^3\lambda_i^2+\frac{d-9}{2}v^2
    +(1-\tau)\lambda_3+\left(1-\frac\tau2\right)v.
  \end{equation}
\end{lemma}
\end{samepage}

\begin{proof}
  Since $H=0$, the six smallest eigenvalues have sum $h$.  By the
  definition of $v$, the first three of them satisfy
  $\lambda_{d-5}+\lambda_{d-4}+\lambda_{d-3}=3v$, so the remaining
  three have sum $h-3v$.  Therefore, using $G\ge0$ on $R$,
  \begin{equation}
    \label{eq:qutrit-G-v}
    G=h-3v-\lambda_1=\lambda_2+\lambda_3-3v\ge0.
  \end{equation}
  Ordering gives $\lambda_3\ge\lambda_{d-5}\ge v$.  It also gives
  $\lambda_{d-2}+\lambda_{d-1}+\lambda_d\le3\lambda_{d-3}\le3v$.
  Since the six smallest eigenvalues have sum $h$, it follows that
  $h\le6v$.  Since $h>0$, we also have $v>0$.  This proves the first three
  inequalities in \eqref{eq:qutrit-basic-compression-ineqs}.

  The middle eigenvalues $\lambda_4,\ldots,\lambda_{d-6}$ lie in
  $[v,\lambda_3]$.  Their sum is $1-2h$, because the three largest and the
  six smallest eigenvalues each have sum $h$.  Hence
  \[
    \sum_{i=4}^{d-6}(\lambda_i-v)
    =(1-2h)-(d-9)v=1-\tau\ge0,
  \]
  with an empty sum when $d=9$, in which case $\tau=1$.  Thus $\tau\le1$.

  For $d-5\le i\le d$, ordering gives $0\le\lambda_i\le\lambda_3\le h/3$.
  We rewrite the sum as
  \begin{equation}
    \label{eq:qutrit-six-smallest-bound}
    \begin{aligned}
      \sum_{i=d-5}^d\lambda_i^2
      &=\lambda_{d-3}h+
      \sum_{i=d-5}^d\lambda_i(\lambda_i-\lambda_{d-3})\\
      &\le\lambda_{d-3}h+
      \frac h3\sum_{i=d-5}^{d-3}(\lambda_i-\lambda_{d-3}).
    \end{aligned}
  \end{equation}
  The last three summands are nonpositive; for the first three, use
  $\lambda_i\le h/3$.
  Since $\sum_{i=d-5}^{d-3}(\lambda_i-\lambda_{d-3})=3v-3\lambda_{d-3}$,
  the last expression in \eqref{eq:qutrit-six-smallest-bound} equals
  $\lambda_{d-3}h+h(v-\lambda_{d-3})=hv$.

  For each middle eigenvalue $\lambda_i$, $4\le i\le d-6$,
  \[
    \lambda_i^2-v^2=(\lambda_i-v)(\lambda_i+v)
    \le(\lambda_3+v)(\lambda_i-v).
  \]
  Summing and using the identity for $1-\tau$ yields
  \[
    \sum_{i=4}^{d-6}\lambda_i^2
    \le(d-9)v^2+(\lambda_3+v)(1-\tau).
  \]
  Equality holds here exactly when every middle eigenvalue equals $v$ or
  $\lambda_3$.  Combining the contributions of the three largest, the middle,
  and the six smallest eigenvalues gives
  \[
    \begin{aligned}
      F&\le\sum_{i=1}^3\lambda_i^2+hv+(d-9)v^2+(\lambda_3+v)(1-\tau)\\
      &=\sum_{i=1}^3\lambda_i^2+\frac{d-9}{2}v^2
      +(1-\tau)\lambda_3+\left(1-\frac\tau2\right)v,
    \end{aligned}
  \]
  where the last equality uses $2h=\tau-(d-9)v$.  This proves
  \eqref{eq:qutrit-compression-bound}.
\end{proof}

For fixed $\tau\in(0,1]$, let $Q_\tau(\lambda_1,\lambda_2,\lambda_3,v)$
denote the right-hand side of \eqref{eq:qutrit-compression-bound}.
This quadratic is convex in the four variables, and strictly convex when
$d>9$.  We maximize it over all quadruples satisfying the equality
\[
  2(\lambda_1+\lambda_2+\lambda_3)+(d-9)v=\tau,
\]
and the inequalities
\[
  \lambda_1\ge\lambda_2\ge\lambda_3\ge v\ge0,\qquad
  h\le6v,\qquad 3v\le\lambda_2+\lambda_3.
\]
These quadruples need not extend to an ordered spectrum.  We optimize over
this larger set, so any upper bound obtained here also holds on the original
$H=0$ face.  Ordering and the equality give $\lambda_1\le\tau/2$.  Also,
$h\le6v$ gives $\tau=2h+(d-9)v\le(d+3)v$, and hence
$v\ge\tau/(d+3)>0$.  Thus the set is a compact polytope for every
$d\ge9$.  Its vertices are the intersections
of the fixed-$\tau$ hyperplane with the extreme rays of the cone defined
by the inequalities.  Convexity therefore reduces the maximum of $Q_\tau$
to these intersections.

To identify the rays, let
\[
  x:=\frac{\lambda_1-\lambda_2}{\lambda_3},\qquad
  y:=\frac{\lambda_2-\lambda_3}{\lambda_3},\qquad
  r:=\frac{v}{\lambda_3}.
\]
These quantities are invariant under positive scaling, and the rays
correspond to points of the polytope defined by
\[
  x,y\ge0,\qquad r\le1,\qquad
  3+2y+x\le6r,\qquad 3r\le2+y.
\]
The face $x=0$ is a quadrilateral, giving rays (i)--(iv) below.
Since the polytope is three-dimensional, at least three linearly independent
constraints are active at each vertex.  If $x>0$, the constraint $x\ge0$
is inactive, so at least three of the remaining four inequalities must be
active.  Since $y=0$ and $r=1$ cannot hold simultaneously, the two
inequalities
\[
  3+2y+x=6r,\qquad 3r=2+y
\]
must both be active, together with either $y=0$ or $r=1$.  The two displayed
equalities give $x=1$ and $y=3r-2$; imposing $y=0$ or $r=1$ then gives
rays (v) and (vi).  We label the six rays by their coordinates
\[
  \left(\frac{\lambda_1}{\lambda_3},
  \frac{\lambda_2}{\lambda_3},
  \frac{v}{\lambda_3}\right)=(1+x+y,1+y,r):
\]
\[
  \begin{array}{lll}
    \text{(i)}\ (1,1,\tfrac12), &
    \text{(ii)}\ (1,1,\tfrac23), &
    \text{(iii)}\ (2,2,1),\\[4pt]
    \text{(iv)}\ (\tfrac52,\tfrac52,1), &
    \text{(v)}\ (2,1,\tfrac23), &
    \text{(vi)}\ (3,2,1).
  \end{array}
\]
The endpoint $\tau=1$ on ray (vi) is realized by
\[
  \boldsymbol\xi_n=\frac1{d+3}(3,2,1^{[d-2]}).
\]
It satisfies $G=H=0$ and has purity $(d+11)/(d+3)^2>M$
for $4\le n\le10$.  Thus $G,H\ge0$ alone do not give the desired
bound in these dimensions.

At vertex (vi), the three active constraints are $r=1$,
$3+2y+x=6r$, and $3r=2+y$.  Each pair of these equalities, together with
the defining inequalities, determines one of the three edges meeting at
(vi).  The other endpoints of these edges are (v), (iii), and (iv).  All
three adjacent vertices lie in the plane $x+3r=3$.  Hence the halfspace
$x+3r\le3$ retains vertices (i)--(v) and excludes (vi).  Since its boundary
contains the three vertices adjacent to (vi), this intersection introduces
no new vertices.
Equivalently, the additional inequality
\begin{equation}
  \label{eq:qutrit-needed-gap}
  \lambda_1-\lambda_2\le3(\lambda_3-v)
\end{equation}
leaves exactly the five rays (i)--(v).

For $n\ge4$, we therefore seek an additional inequality, valid for APPT
spectra, that implies \eqref{eq:qutrit-needed-gap} on $H=0$.
We first establish the purity bound under this condition.

\begin{lemma}
  \label{lem:qutrit-H-face}
  For $n\ge3$, use the notation of Lemma~\ref{lem:qutrit-H-compression}
  on $R\cap\{H=0\}$.  Then
  \begin{equation}
    \label{eq:qutrit-face-bound}
    \max_{\substack{\boldsymbol\lambda\in R,\ H=0\\
        \lambda_1-\lambda_2\le3(\lambda_3-v)}}F(\boldsymbol\lambda)
    =\begin{cases}
      \dfrac{d+5}{(d+1)^2}, & n=3,4,\\[6pt]
      \dfrac{2d+21}{2(d+3)^2}, & 5\le n\le8,\\[6pt]
      M_{\mathrm{cen}},                   & n\ge9.
    \end{cases}
  \end{equation}
  This bound also holds on the $H=0$ part of either face of $R$ defined by
  $\lambda_1=\lambda_2$ or $\lambda_2=\lambda_3$.
  We call these the \emph{repeated-pair faces}.

  For $n\ge4$, the right-hand side of
  \eqref{eq:qutrit-face-bound} is at most $M$, and it is strictly below
  $M$ for $4\le n\le8$.  If $n=3$, the whole $H=0$ face of $R$ satisfies
  $F\le7/50<M_{\mathrm{one}}$.
\end{lemma}

\begin{proof}
  Imposing $\lambda_1-\lambda_2\le3(\lambda_3-v)$ leaves rays (i)--(v).
  For each fixed $\tau$, convexity of $Q_\tau$ reduces its maximum to the
  five ray intersections.  We then vary the scale along each ray, so
  $\tau=2h+(d-9)v$ is no longer fixed.

  \smallskip \noindent\emph{Bounds along the five rays.}
  On rays (i)--(iii), the values of $Q_\tau$ satisfy
  \[
    \begin{aligned}
      \text{(i)}:\quad&\frac32\lambda_3-\frac d2\lambda_3^2
      =M_{\mathrm{cen}}-\frac d2\left(\lambda_3-\frac{3}{2d}\right)^2
      \le M_{\mathrm{cen}},\\[3pt]
      \text{(ii)}:\quad&\frac53\lambda_3-\frac{2d-3}{3}\lambda_3^2
      \le\frac{25}{12(2d-3)},\\[3pt]
      \text{(iii)}:\quad&2\lambda_3-(d-3)\lambda_3^2\le\frac{d+5}{(d+1)^2}.
    \end{aligned}
  \]
  The bound for (ii) follows by completing the square.  For (iii),
  $\tau\le1$ gives $\lambda_3\le1/(d+1)$, and the quadratic is increasing on
  that interval.  On rays (iv) and (v), $\tau=(d+3)v$.  Writing
  \[
    q_d(v):=\frac52v-\frac32(d-2)v^2,
    \qquad 0\le v\le\frac1{d+3},
  \]
  the corresponding values of $Q_\tau$ are
  \[
    \begin{aligned}
      \text{(iv)}:\quad&q_d(v)-\frac12v(1-\tau),\\[3pt]
      \text{(v)}:\quad&q_d(v).
    \end{aligned}
  \]
  Since $\tau\le1$, both expressions are bounded above by $q_d(v)$.
  On ray (iv), the endpoint $v=1/(d+3)$ has $\tau=1$ and is realized by
  \[
    \boldsymbol\eta_n:=\frac1{2(d+3)}(5,5,2^{[d-2]}),
    \qquad
    B_\eta:=F(\boldsymbol\eta_n)=\frac{2d+21}{2(d+3)^2}.
  \]
  For $n\le8$, $q_d$ is increasing on $[0,1/(d+3)]$, so
  $q_d(v)\le q_d(1/(d+3))=B_\eta$.
  For $n\ge9$, completing the square gives
  \[
    q_d(v)\le\frac{25}{24(d-2)}\le M_{\mathrm{cen}},\qquad
    M_{\mathrm{cen}}-\frac{25}{24(d-2)}
    =\frac{d-27}{12d(d-2)}\ge0.
  \]
  \Needspace{5\baselineskip}
  \smallskip \noindent\emph{Comparison of the resulting bounds.}
  Since $d=3n$,
  \[
    B_\eta-\frac{d+5}{(d+1)^2}=\frac{9n^2-34n-23}
    {6(n+1)^2(3n+1)^2},\qquad
    M_{\mathrm{cen}}-B_\eta=\frac{(n-9)(n-1)}{24n(n+1)^2}.
  \]
  For $n=3,4$, these identities give
  $M_{\mathrm{cen}}<B_\eta<(d+5)/(d+1)^2$.
  For $5\le n\le8$, $B_\eta$ is the largest of the three bounds;
  for $n\ge9$, $M_{\mathrm{cen}}$ is the largest.  This gives the three
  branches in \eqref{eq:qutrit-face-bound}.
  The bound for ray (ii) is never the largest, since
  \[
    \frac{d+5}{(d+1)^2}-\frac{25}{12(2d-3)}
    =\frac{(d-9)(25-d)+20}{12(d+1)^2(2d-3)}>0
    \qquad(9\le d\le24),
  \]
  whereas for $n\ge9$ it is strictly below the common bound just obtained:
  \[
    \frac{25}{12(2d-3)}<\frac{25}{24(d-2)}\le M_{\mathrm{cen}}.
  \]
  For $n=4$, the value $\frac{d+5}{(d+1)^2}$ is strictly below
  $M_{\mathrm{one}}$ by the edge bound \eqref{eq:qutrit-G-edge-bound}.  For
  $5\le n\le8$,
  \[
    M_{\mathrm{one}}-B_\eta=\frac{4+(8-n)(3n+2)}
    {6(n+1)^2(3n+2)^2}>0.
  \]
  Hence the right-hand side of \eqref{eq:qutrit-face-bound} is at most
  $M$ for every $n\ge4$, and is strictly below $M$ for $4\le n\le8$.
  Each branch is attained by the corresponding spectrum below:
  \[
    \begin{array}{c@{\qquad}c}
      \text{Range} & \text{An attaining ordered spectrum}\\[4pt]
      n=3,4 & \dfrac{(2,2,1^{[d-3]},0)}{d+1}\\[8pt]
      5\le n\le8 & \dfrac{(5,5,2^{[d-2]})}{2(d+3)}\\[8pt]
      n\ge9 & \dfrac{(2^{[n]},1^{[2n]})}{4n}.
    \end{array}
  \]
  This proves \eqref{eq:qutrit-face-bound}.

  For $n\ge9$, among the five rays, equality $Q_\tau=M_{\mathrm{cen}}$
  occurs only on ray (i), with $\lambda_3=3/(2d)$, except that rays (iv) and (v)
  also attain this value when $n=9$ and $\tau=1$.

  \smallskip \noindent\emph{Repeated-pair faces.}
  The gap condition \eqref{eq:qutrit-needed-gap} holds automatically on
  either repeated-pair face.  If $\lambda_1=\lambda_2$, it follows from
  $\lambda_3\ge v$.  If $\lambda_2=\lambda_3$, then
  \[
    3(\lambda_3-v)-(\lambda_1-\lambda_2)=(6v-h)+3G\ge0,
  \]
  since $h\le6v$ and $G\ge0$.
  Thus \eqref{eq:qutrit-face-bound} applies to both faces.

  \smallskip \noindent\emph{The case $n=3$.}
  Here $d=9$ and $\tau=1$ by Lemma~\ref{lem:qutrit-H-compression}.
  Without \eqref{eq:qutrit-needed-gap}, the only additional direction is
  ray (vi).  Its intersection with $\tau=1$ has $\lambda_3=1/12$ and
  $Q_1=2\lambda_3-4\lambda_3^2=5/36<7/50$.
  Thus the whole $H=0$ face satisfies $F\le7/50<M_{\mathrm{one}}$.
\end{proof}

\subsection{An additional inequality from the Hildebrand criterion}

For $n\ge4$, we impose an additional inequality that implies
$\lambda_1-\lambda_2<3(\lambda_3-v)$ on $H=0$ and excludes simultaneous
equality $G=H=0$.  The first property allows us to apply the preceding
purity bound; the second limits the number of active added inequalities
at the remaining vertices.

Direct substitution shows that $K(\boldsymbol\xi_n)$ has the negative
eigenvalue $(2-\sqrt5)/(d+3)$ with eigenvector $(\sqrt5,2,1)^\top$.
This vector supplies the required test for $n\ge4$.
For $n=3$, however, it does not exclude $\boldsymbol\nu_7$, since
$7\ell_{(\sqrt5,2,1)^\top}(\boldsymbol\nu_7)=5-2\sqrt5>0$.
We therefore use $(2,1,0)^\top$ in that dimension and define
\[
  W_n:=
  \begin{cases}
    \ell_{(2,1,0)^\top},       & n=3,\\[3pt]
    \ell_{(\sqrt5,2,1)^\top}, & n\ge4.
  \end{cases}
\]
For every $n\ge3$, put
\[
  \CT_n:=\{\boldsymbol\lambda\in R:W_n(\boldsymbol\lambda)\ge0\}.
\]
Since $W_n\ge0$ on $\CA_{3,n}$,
\begin{equation}
  \label{eq:qutrit-outer-inclusion}
  \CA_{3,n}\subseteq\CT_n\subseteq R.
\end{equation}
When $n=3$, direct substitution gives
\begin{equation}
  \label{eq:qutrit-n3-cut}
  W_3=\lambda_7+2\lambda_8+4\lambda_9-2\lambda_1
  =2G-\lambda_7+2\lambda_9.
\end{equation}

For $n\ge4$, both required properties follow from the rearrangement
estimate below.
Retain the definition of $v$ from Lemma~\ref{lem:qutrit-H-compression}
and put $T:=\lambda_{d-2}+\lambda_{d-1}+\lambda_d$.
Direct expansion gives
\[
  \begin{aligned}
    W_n
    ={}&\lambda_{d-5}+2\lambda_{d-4}+\sqrt5\,\lambda_{d-3}\\
    &+4\lambda_{d-2}+2\sqrt5\,\lambda_{d-1}+5\lambda_d
    -2\sqrt5\,\lambda_1-\sqrt5\,\lambda_2-2\lambda_3.
  \end{aligned}
\]
The coefficient triples $(1,2,\sqrt5)$ and $(4,2\sqrt5,5)$ are increasing,
whereas each corresponding triple of eigenvalues is nonincreasing.
Replacing each triple of coefficients by its average can therefore only
increase the sum.  Thus, for every $\boldsymbol\lambda\in\CO_d$,
\begin{equation}
  \label{eq:qutrit-two-triple-estimate}
  W_n\le(\sqrt5+3)v+\frac{9+2\sqrt5}{3}T
  -2\sqrt5\,\lambda_1-\sqrt5\,\lambda_2-2\lambda_3.
\end{equation}

\begin{samepage}
\begin{lemma}
  \label{lem:qutrit-gap-bound}
  Let $n\ge4$ and $\boldsymbol\lambda\in\CO_d$ satisfy $H=0$ and $W_n\ge0$.
  Then
  \begin{equation}
    \label{eq:qutrit-quantitative-gap}
    \lambda_1-\lambda_2
    \le3(\lambda_3-v)-(\sqrt5-2)^2\lambda_3.
  \end{equation}
  In particular, $\lambda_1-\lambda_2<3(\lambda_3-v)$.
\end{lemma}
\end{samepage}

\begin{proof}
  The equality $H=0$ and ordering give
  \[
    h=3v+T,\qquad 0\le T\le3v\le3\lambda_3.
  \]
  Also $0<h=\sum_{i=d-5}^d\lambda_i\le6\lambda_3$, so $\lambda_3>0$.
  Substituting $\lambda_2=3v+T-\lambda_1-\lambda_3$ into
  \eqref{eq:qutrit-two-triple-estimate} yields
  \[
    \begin{aligned}
      0\le W_n
      &\le-\sqrt5\,\lambda_1+(\sqrt5-2)\lambda_3
      -(2\sqrt5-3)v+\frac{9-\sqrt5}{3}T\\
      &\le-\sqrt5\,\lambda_1+(\sqrt5-2)\lambda_3+(4-\sqrt5)T,
    \end{aligned}
  \]
  where $T\le3v$ gives the second inequality.
  Solving for $\lambda_1$ and using $H=0$ again, we obtain
  \[
    \begin{aligned}
      \lambda_1-\lambda_2-3(\lambda_3-v)
      &=2\lambda_1-T-2\lambda_3\\
      &\le\frac{8-3\sqrt5}{\sqrt5}\,T-\frac4{\sqrt5}\lambda_3\\
      &\le(4\sqrt5-9)\lambda_3
      =-(\sqrt5-2)^2\lambda_3.
    \end{aligned}
  \]
  The last inequality uses $T\le3\lambda_3$ and $8-3\sqrt5>0$.
  Since $\lambda_3>0$, the correction term is positive, proving the
  strict conclusion.
\end{proof}

\begin{lemma}
  \label{lem:qutrit-third-cut}
  Let $n\ge4$.  If $\boldsymbol\lambda\in R$ satisfies $G=H=0$, then
  \[
    W_n(\boldsymbol\lambda)
    \le-\frac32(\sqrt5-2)\lambda_1<0.
  \]
\end{lemma}

\begin{proof}
  On $G=H=0$, we have
  \[
    \lambda_1=T>0,\qquad \lambda_2+\lambda_3=3v\ge T.
  \]
  Since $\lambda_2\ge\lambda_3$,
  \[
    \sqrt5\,\lambda_2+2\lambda_3
    \ge\frac{\sqrt5+2}{2}(\lambda_2+\lambda_3).
  \]
  Applying \eqref{eq:qutrit-two-triple-estimate} and using $3v\ge T=\lambda_1$
  gives
  \[
    W_n\le-\frac{\sqrt5}{2}v+\frac{(\sqrt5-2)^2}{3}T
    \le-\frac32(\sqrt5-2)\lambda_1<0.\qedhere
  \]
\end{proof}

\subsection{The remaining vertices}

For $n\ge4$, Lemma~\ref{lem:qutrit-gap-bound} gives the strict form of
\eqref{eq:qutrit-needed-gap} on $\CT_n\cap\{H=0\}$, so
Lemma~\ref{lem:qutrit-H-face} gives $F\le M$ there.
Lemma~\ref{lem:qutrit-third-cut} excludes simultaneous equality $G=H=0$.
At most two of $G\ge0$, $H\ge0$, and $W_n\ge0$ can therefore be active.
Applying \eqref{eq:face-dimension-bound} relative to $\CO_d$ gives at most
three distinct positive eigenvalues at every vertex of $\CT_n$.

\begin{lemma}
  \label{lem:qutrit-residual-classification}
  Let $n\ge4$, and let $\boldsymbol\lambda$ be a vertex of $\CT_n$ with
  $\lambda_1>\lambda_2>\lambda_3$.  Then $\boldsymbol\lambda$ is one of
  \[
    \frac{1}{d+1+\beta_0}
    \bigl(3,\beta_0,1^{[d-2]}\bigr),
    \qquad
    \frac{1}{d-1+\beta_1}
    \bigl(2,\beta_1,1^{[d-3]},0\bigr),
  \]
  where
  \[
    \beta_0:=2\sqrt5-3,\qquad \beta_1:=\sqrt5-1.
  \]
\end{lemma}

\begin{proof}
  Since $\lambda_1>\lambda_2>\lambda_3$, the rank observation in
  Subsection~\ref{sec:qutrit-prelim} gives rank $d-1$ or $d$.
  The bound of three distinct positive eigenvalues therefore gives
  \[
    \boldsymbol\lambda
    =\bigl(\lambda_1,\lambda_2,\lambda_3^{[d-\zeta-2]},0^{[\zeta]}\bigr),
    \qquad \lambda_1>\lambda_2>\lambda_3>0,
  \]
  where $\zeta\in\{0,1\}$ counts the zero eigenvalues.
  Since $v=\lambda_3$ and $\lambda_1>\lambda_2$,
  Lemma~\ref{lem:qutrit-gap-bound} excludes $H=0$.  Thus $H>0$.

  The spectrum lies in the relative interior of the triangle
  $[\boldsymbol\nu_1,\boldsymbol\nu_2,\boldsymbol\nu_{d-\zeta}]$.
  Since it is a vertex of $\CT_n$, two independent added inequalities
  must be active.  These are $G=W_n=0$, because $H>0$.
  The first equality gives $\lambda_1=(3-\zeta)\lambda_3$, after which
  \[
    W_n=\sqrt5\bigl(\beta_\zeta\lambda_3-\lambda_2\bigr).
  \]
  Hence $\lambda_2=\beta_\zeta\lambda_3$, and normalization gives the
  stated spectra.
\end{proof}

\begin{lemma}
  \label{lem:qutrit-residual-purity}
  Let $n\ge4$, and let $\boldsymbol\lambda$ be a vertex of $\CT_n$ with
  $\lambda_1>\lambda_2>\lambda_3$.  Then $F(\boldsymbol\lambda)<M$.
\end{lemma}

\begin{proof}
  \emph{Spectrum of rank $d-1$.}
  By Lemma~\ref{lem:qutrit-residual-classification}, the spectrum lies in
  the relative interior of
  $[\boldsymbol\nu_1,\boldsymbol\nu_2,\boldsymbol\nu_{d-1}]$ and on $G=0$.
  The intersection with $G=0$ is the segment joining
  \[
    \frac1d(2,1^{[d-2]},0)
    \qquad\text{and}\qquad
    \frac1{d+1}(2,2,1^{[d-3]},0).
  \]
  These are the $G$-edge points of rank $d-1$ with $i=1,2$, so both have
  purity strictly below $M_{\mathrm{one}}$ by
  Proposition~\ref{prop:qutrit-G-edges}.  Convexity gives the same strict
  bound throughout the segment, hence
  $F(\boldsymbol\lambda)<M_{\mathrm{one}}\le M$.

  \smallskip \noindent\emph{Full-rank spectrum.}
  Put $\beta:=\beta_0=2\sqrt5-3=\lambda_2/\lambda_3$,
  so $1<\beta<3/2$.
  Since $\lambda_3=(d+1+\beta)^{-1}$, we have
  \[
    \boldsymbol\lambda
    =\bigl(3\lambda_3,\,1-(d+1)\lambda_3,\,\lambda_3^{[d-2]}\bigr),
    \qquad \frac{2}{2d+5}<\lambda_3<\frac1{d+2}.
  \]
  This expression is affine in $\lambda_3$, so $\boldsymbol\lambda$
  lies strictly inside the segment joining
  $\boldsymbol\lambda^{\mathrm{one}}$ and
  \[
    \boldsymbol\omega:=\frac1{2d+5}(6,3,2^{[d-2]}),
    \qquad F(\boldsymbol\omega)=\frac{4d+37}{(2d+5)^2}.
  \]

  For $4\le n\le7$, we have $12\le d\le21$, and
  \[
    M_{\mathrm{one}}-F(\boldsymbol\omega)
    =\frac{52+d(21-d)}{(d+2)^2(2d+5)^2}>0.
  \]
  Convexity therefore gives $F(\boldsymbol\lambda)<M_{\mathrm{one}}$.
  For $n\ge9$, the first endpoint has purity
  $M_{\mathrm{one}}<M_{\mathrm{cen}}$, while
  \[
    M_{\mathrm{cen}}-F(\boldsymbol\omega)
    =\frac{4(d-27)(d-2)+9}{8d(2d+5)^2}>0.
  \]
  Convexity again gives $F(\boldsymbol\lambda)<M_{\mathrm{cen}}$.

  Finally, when $n=8$ and $d=24$, direct subtraction gives
  \[
    F(\boldsymbol\lambda)-M_{\mathrm{one}}
    =\frac{(\beta-1)(161\beta-239)}{169(\beta+25)^2}<0,
  \]
  since $1<\beta=2\sqrt5-3<239/161$.
\end{proof}

\subsection{The maximum and equality cases}

\begin{lemma}
  \label{lem:qutrit-outer-maximum}
  For every $n\ge3$, the purity has a unique maximizer on
  $\CT_n$.  It is the first spectrum in
  \eqref{eq:qutrit-maximizers} for $3\le n\le8$, and the second for
  $n\ge9$.  The corresponding maximum is the appropriate branch of
  \eqref{eq:qutrit-maximum}.
\end{lemma}

\begin{proof}
  Both candidate spectra belong to $\CA_{3,n}\subseteq\CT_n$, and
  \eqref{eq:qutrit-two-candidate-gap} identifies the one with purity $M$
  in each dimension range.  It remains to prove $F\le M$ on $\CT_n$ and
  identify all equality cases.

  \smallskip \noindent\emph{The remaining vertices for $n\ge4$.}
  Ordering makes each repeated-pair set $\lambda_1=\lambda_2$ or
  $\lambda_2=\lambda_3$ a face of $R$, so its vertices are vertices of $R$.
  Those with $H=0$ satisfy the bound in
  Lemma~\ref{lem:qutrit-H-face}, while those with $H>0$ satisfy the
  preliminary simplex and $G$-edge bounds by
  Lemma~\ref{lem:active-face-dimension}.
  Convexity therefore gives $F\le M$ on both entire faces of $R$, hence
  on their intersections with $\CT_n$.
  Every remaining vertex of $\CT_n$ satisfies
  $\lambda_1>\lambda_2>\lambda_3$ and has purity strictly below $M$ by
  Lemma~\ref{lem:qutrit-residual-purity}.  Thus $F\le M$ throughout
  $\CT_n$, and every maximizer lies on a repeated-pair face.

  \smallskip \noindent\emph{Equality for $n\ge4$.}
  Since each entire repeated-pair face of $R$ has purity at most $M$,
  strict convexity forces any point attaining $M$ in its intersection with
  $\CT_n$ to be a vertex of that face of $R$.
  When $H>0$, the preliminary bounds allow equality only at
  $\boldsymbol\lambda^{\mathrm{one}}$, and only for $4\le n\le8$.
  On $H=0$, Lemma~\ref{lem:qutrit-H-face} is already strict for
  $4\le n\le8$.  For $n\ge9$, the quadruple associated with a maximizing
  spectrum belongs to the fixed-$\tau$ polytope used in that lemma, and
  \[
    M_{\mathrm{cen}}=F(\boldsymbol\lambda)
    \le Q_\tau(\lambda_1,\lambda_2,\lambda_3,v)\le M_{\mathrm{cen}}.
  \]
  Both inequalities are equalities.  The strict convexity of $Q_\tau$ at
  fixed $\tau$ (since $d>9$) forces the quadruple to be a vertex of that
  polytope, hence to lie on one of rays (i)--(v).  The ray equality conditions
  in Lemma~\ref{lem:qutrit-H-face} leave only ray (i), apart from rays
  (iv) and (v) when $n=9$.  These satisfy
  $\lambda_1-\lambda_2=3(\lambda_3-v)$ and are excluded by
  Lemma~\ref{lem:qutrit-gap-bound}.  Thus only ray (i) remains.
  Equality on this ray gives
  $\lambda_1=\lambda_2=\lambda_3=1/(2n)$ and $v=1/(4n)$, so $h=6v$.
  The upper and lower triples of the six smallest eigenvalues therefore
  have the same sum $3v$.  Ordering forces
  $\lambda_{d-5}=\cdots=\lambda_d=v$.  Equality in the
  middle-eigenvalue estimate of Lemma~\ref{lem:qutrit-H-compression} gives
  $\lambda_i\in\{v,\lambda_3\}$ for $4\le i\le d-6$.
  Thus every eigenvalue equals $v$ or $2v$.  If $k$ denotes the total
  number of eigenvalues equal to $2v$, normalization gives
  $1=(d+k)v$.  Since $d=3n$ and $v=1/(4n)$, we obtain $k=n$, and hence
  the central spectrum.

  \Needspace{8\baselineskip}
  \smallskip \noindent\emph{The case $n=3$.}
  Lemma~\ref{lem:qutrit-H-face} bounds the whole $H=0$ face by
  $7/50<M_{\mathrm{one}}$.  Together with the preliminary simplex and
  $G$-edge bounds, this leaves $\boldsymbol\nu_7$ as the sole vertex of
  $R$ with purity above $M_{\mathrm{one}}$.  The inequality $W_3\ge0$
  excludes it, since $W_3(\boldsymbol\nu_7)=-1/7$.
  By Lemma~\ref{lem:active-face-dimension}, intersecting $R$ with this
  halfspace creates new vertices only on edges of $R$.  On an edge not
  through $\boldsymbol\nu_7$, both endpoint purities are at most
  $M_{\mathrm{one}}$, so strict convexity gives $F<M_{\mathrm{one}}$ at every
  interior point.

  Since $H(\boldsymbol\nu_7)=1/7>0$, the inequality $H\ge0$ is strict in the
  relative interior of every edge through $\boldsymbol\nu_7$.  The
  face-dimension bound \eqref{eq:face-dimension-bound} and the signs of $G$
  therefore give two possibilities: a simplex edge
  $[\boldsymbol\nu_7,\boldsymbol\nu_k]$, $k=8,9$, or an edge contained
  in $G=0$ in a triangle
  $[\boldsymbol\nu_i,\boldsymbol\nu_7,\boldsymbol\nu_k]$,
  $1\le i\le6$, $k=8,9$.

  The intersections of the two simplex edges with $W_3=0$ are
  \[
    \frac1{15}(2^{[7]},1,0),\qquad
    \frac1{44}(6^{[7]},1,1).
  \]
  Their purities are $29/225$ and $127/968$, respectively, both strictly
  below $3/22$.

  On $G=W_3=0$, identity~\eqref{eq:qutrit-n3-cut} gives
  $\lambda_7=2\lambda_9$.  Thus $k=9$, since $k=8$ would have
  $\lambda_9=0<\lambda_7$.  A spectrum in the triangle
  $[\boldsymbol\nu_i,\boldsymbol\nu_7,\boldsymbol\nu_9]$ has its first
  $i$ entries equal to $\lambda_1$, its next $7-i$ equal to $\lambda_7$,
  and its last two equal to $\lambda_9$.  Since $G=0$ also gives
  $\lambda_1=\lambda_7+2\lambda_9=4\lambda_9$, normalization yields
  \[
    \boldsymbol\mu_i
    =\frac1{2i+16}(4^{[i]},2^{[7-i]},1,1),
    \qquad 1\le i\le6.
  \]
  For every $1\le i\le6$,
  \[
    F(\boldsymbol\mu_i)
    =\frac3{22}-\frac{3(i-3)^2}{22(i+8)^2}\le\frac3{22}.
  \]
  Thus every new vertex on an edge through $\boldsymbol\nu_7$
  has purity at most
  $3/22<17/121=M_{\mathrm{one}}$.  Together with the preliminary bounds and
  strict convexity, this leaves $\boldsymbol\lambda^{\mathrm{one}}$ as the
  unique maximizer.
\end{proof}

\raggedbottom
\begin{samepage}
\begin{proof}[Proof of Theorem~\ref{thm:qutrit}]
  Both candidate spectra are APPT by the direct verification in
  Subsection~\ref{sec:qutrit-criterion}.  The inclusion
  \eqref{eq:qutrit-outer-inclusion} and
  Lemma~\ref{lem:qutrit-outer-maximum} therefore give the APPT maximum and
  the unique maximizing ordered spectrum.  Unitary invariance gives exactly
  the stated APPT maximizers.

  It remains only to pass to absolute separability.  Every absolutely
  separable state is APPT, so the same value is an upper bound.
  By \cite[Theorem~3]{AV}, the condition
  $\lambda_{\min}\ge1/(d+2)$ is sufficient for absolute separability;
  $\boldsymbol\lambda^{\mathrm{one}}$ meets it with equality.
  By \cite[Theorem~1]{Kondra}, the condition
  $\lambda_{\max}/\lambda_{\min}\le2$ is also sufficient on
  $\BC^3\otimes\BC^n$; $\boldsymbol\lambda^{\mathrm{cen}}$ meets it with
  equality.  Thus the relevant
  candidate attains the same maximum in the smaller set.  Any absolutely
  separable maximizer is also an APPT maximizer, so the uniqueness and
  equality statements transfer unchanged.
\end{proof}
\end{samepage}

\section{Quantitative consequences}
\label{sec:quantitative-consequence}

\smallskip \noindent\textbf{Geometric interpretation.}
The purity maximum has an enclosing-ball interpretation; see, for
example, \cite{DK}.  For the Hilbert--Schmidt norm
$\|X\|_{\mathrm{HS}}:=\sqrt{\operatorname{Tr}(X^\dagger X)}$,
every state on a $d$-dimensional space satisfies
\[
  \left\|\rho-\frac{I_d}{d}\right\|_{\mathrm{HS}}^2
  =\operatorname{Tr}(\rho^2)-\frac1d.
\]
Thus maximizing purity is equivalent to finding the states farthest from the
maximally mixed state.  Since Theorem~\ref{thm:qutrit} gives the same maximum
and maximizing orbit for the absolutely separable and APPT sets, their exact
maximum distances coincide:
\begin{equation}
  \label{eq:qutrit-HS-radius}
  R^{\mathrm{AS}}_{3,n}=R^{\mathrm{APPT}}_{3,n}=:R_{3,n}
  =\begin{cases}
    \dfrac{2}{3n+2}\sqrt{\dfrac{3n-1}{3n}}, & 3\le n\le8,\\[8pt]
    \dfrac1{\sqrt{24n}}, & n\ge9.
  \end{cases}
\end{equation}

\smallskip \noindent\textbf{Partial-transpose eigenvalues.}
For a state $\rho$ on $\BC^3\otimes\BC^n$, define
\[
  q(\rho):=\min_{U\in\mathcal U(d)}
  \lambda_{\min}\!\left((U\rho U^\dagger)^\Gamma\right),
\]
where $\mathcal U(d)$ denotes the group of unitary operators on $\BC^d$.
The minimum exists because $\mathcal U(d)$ is compact and the objective is
continuous.  Thus $\rho$ is APPT exactly when $q(\rho)\ge0$.  Because the
minimization is over the full global unitary orbit, $q(\rho)$ depends only
on the spectrum.  When $q(\rho)<0$, the quantity $-q(\rho)$ measures how
far below zero the smallest partial-transpose eigenvalue can be lowered by a
global unitary.

\begin{samepage}
\begin{lemma}
  \label{lem:qutrit-min-pt-matrices}
  If $\boldsymbol\lambda$ is the ordered spectrum of $\rho$, then
  \begin{equation}
    \label{eq:qutrit-min-pt-matrices}
    2q(\rho)=
    \min\left\{
      \lambda_{\min}\!\left(K(\boldsymbol\lambda)\right),
      \lambda_{\min}\!\left(\widetilde K(\boldsymbol\lambda)\right)
    \right\}.
  \end{equation}
  Moreover, if $\rho\ne I_d/d$, then $q(\rho)\le\lambda_d<1/d$, and
  \[
    \frac{\rho-q(\rho)I_d}{1-dq(\rho)}
  \]
  is an APPT state.
\end{lemma}
\end{samepage}

\begin{proof}
  For the maximally mixed state, $q(\rho)=1/d$ and
  $K=\widetilde K=2I_3/d$, so the formula holds.
  Otherwise, a global unitary can make $\rho$ diagonal in a product basis,
  giving $q(\rho)\le\lambda_d<1/d$.
  For every $s\le\lambda_d$, the operator
  $(\rho-sI_d)/(1-ds)$ is a density matrix.
  Its APPT property is equivalent to $q(\rho)\ge s$ by definition,
  and the qutrit criterion therefore gives
  \[
    q(\rho)\ge s
    \quad\Longleftrightarrow\quad
    K(\boldsymbol\lambda)-2sI_3\succeq0
    \quad\text{and}\quad
    \widetilde K(\boldsymbol\lambda)-2sI_3\succeq0.
  \]
  The right-hand side holds exactly when $s$ is at most half the smaller
  of the two minimum eigenvalues.  This endpoint is at most $\lambda_d$,
  since the first diagonal entry of $K$ is $2\lambda_d$.
  Hence \eqref{eq:qutrit-min-pt-matrices} follows.
  Taking $s=q(\rho)$ proves the final assertion.
\end{proof}

Theorem~\ref{thm:qutrit} gives a bound depending only on the purity.

\begin{corollary}
  \label{cor:qutrit-min-pt}
  Let $P=\operatorname{Tr}(\rho^2)$, and let $M$ be the value in
  \eqref{eq:qutrit-maximum}.  Then
  \begin{equation}
    \label{eq:qutrit-min-pt-bound}
    q(\rho)\le
    \frac1d\left(
      1-\sqrt{\frac{P-1/d}{M-1/d}}
    \right).
  \end{equation}
  Let $\boldsymbol\lambda^*$ be the maximizing spectrum from
  Theorem~\ref{thm:qutrit}.  Equality holds exactly when the ordered spectrum
  is a nonnegative vector of the form
  \begin{equation}
    \label{eq:qutrit-min-pt-equality}
    \boldsymbol u+t(\boldsymbol\lambda^*-\boldsymbol u),
    \qquad
    \boldsymbol u=(1/d)^{[d]},
    \qquad
    t=\sqrt{\frac{P-1/d}{M-1/d}}.
  \end{equation}
  Consequently, equality occurs
  for every $P\in[1/d,1]$ when $3\le n\le8$, and exactly for
  $P\in[1/d,1/n]$ when $n\ge9$.  For $n\ge9$ and $P>1/n$, the inequality
  in \eqref{eq:qutrit-min-pt-bound} is strict.
\end{corollary}

\begin{proof}
  For the maximally mixed state, $P=1/d$ and $q(\rho)=1/d$, so the claim
  holds.  Otherwise, write $q=q(\rho)$.  By
  Lemma~\ref{lem:qutrit-min-pt-matrices},
  \[
    \sigma:=\frac{\rho-qI_d}{1-dq}
  \]
  is an APPT state.  Theorem~\ref{thm:qutrit} gives
  \[
    \operatorname{Tr}(\sigma^2)-\frac1d
    =\frac{P-1/d}{(1-dq)^2}
    \le M-\frac1d,
  \]
  which is equivalent to \eqref{eq:qutrit-min-pt-bound}.

  Equality holds only if $\sigma$ is an APPT maximizer, so by
  Theorem~\ref{thm:qutrit} its ordered spectrum is
  $\boldsymbol\lambda^*$.  Since
  $\rho=(1-t)I_d/d+t\sigma$ with $t=1-dq$, this gives
  \eqref{eq:qutrit-min-pt-equality} and the displayed value of $t$.
  Conversely, let $\rho_*$ have spectrum $\boldsymbol\lambda^*$.
  The candidate matrices in Subsection~\ref{sec:qutrit-criterion} are
  positive semidefinite and singular, so
  Lemma~\ref{lem:qutrit-min-pt-matrices} gives $q(\rho_*)=0$.
  Therefore, for
  $\rho_t=(1-t)I_d/d+t\rho_*$ with $t\ge0$ and $\rho_t\succeq0$,
  \[
    q(\rho_t)=\frac{1-t}{d}+tq(\rho_*)=\frac{1-t}{d},
    \qquad
    \operatorname{Tr}(\rho_t^2)-\frac1d
    =t^2\left(M-\frac1d\right),
  \]
  so every nonnegative spectrum in \eqref{eq:qutrit-min-pt-equality}
  attains equality.  For $3\le n\le8$, nonnegativity is equivalent to
  $0\le t\le(d+2)/2$, whose upper endpoint is a pure spectrum.  For
  $n\ge9$, it is equivalent to $0\le t\le4$, whose upper endpoint is
  $(1/n)^{[n]},0^{[2n]}$.  These ranges give the stated purity intervals.
\end{proof}

\smallskip \noindent\textbf{Depolarizing noise.}
For $0\le p\le1$, let
\[
  \mathcal E_p(\rho):=(1-p)\rho+\frac{p}{d}I_d.
\]
Global depolarization commutes with unitary conjugation and partial
transposition, and therefore
\[
  q\!\left(\mathcal E_p(\rho)\right)
  =(1-p)q(\rho)+\frac{p}{d}.
\]
Define the minimum amounts of depolarizing noise by
\[
  \begin{aligned}
    p_{\rm AS}(\rho)
    &:=\inf\{p\in[0,1]:\mathcal E_p(\rho)
      \text{ is absolutely separable}\},\\
    p_{\rm APPT}(\rho)
    &:=\inf\{p\in[0,1]:\mathcal E_p(\rho)
      \text{ is APPT}\}.
  \end{aligned}
\]
Both infima are attained, since the two classes are closed and contain
$I_d/d$, and $p_{\rm AS}(\rho)\ge p_{\rm APPT}(\rho)$.
If $q(\rho)\ge0$, then $p_{\rm APPT}(\rho)=0$.
If $q(\rho)<0$, the preceding identity gives
\begin{equation}
  \label{eq:qutrit-depolarizing-strength}
  p_{\rm APPT}(\rho)
  =\frac{-d\,q(\rho)}{1-d\,q(\rho)}.
\end{equation}

\begin{corollary}
  \label{cor:qutrit-depolarizing-purity}
  Let $P=\operatorname{Tr}(\rho^2)$.
  If $P>M$, then
  \begin{equation}
    \label{eq:qutrit-depolarizing-purity-bound}
    p_{\rm AS}(\rho)\ge p_{\rm APPT}(\rho)
    \ge 1-\sqrt{\frac{M-1/d}{P-1/d}}.
  \end{equation}
  For $P>M$, either quantity equals the displayed lower bound if and only
  if the ordered spectrum belongs to the equality family
  \eqref{eq:qutrit-min-pt-equality}; in that case both quantities equal it.

  Thus the lower bound is the exact minimum of each quantity over states
  of purity $P$ for $3\le n\le8$ and $M<P\le1$, and for $n\ge9$ and
  $M<P\le1/n$.
  For $n\ge9$ and $P>1/n$, both quantities are strictly greater than the
  displayed lower bound.
  For $1/d\le P\le M$, the minimum of each quantity over states of
  purity $P$ is zero.
\end{corollary}

\begin{proof}
  The identity
  \[
    \operatorname{Tr}\!\left(\mathcal E_p(\rho)^2\right)
    =\frac1d+(1-p)^2\left(P-\frac1d\right)
  \]
  and Theorem~\ref{thm:qutrit} give the stated lower bound for either
  quantity when $P>M$.

  If either quantity equals the lower bound, the corresponding
  depolarized state has purity $M$.  By Theorem~\ref{thm:qutrit}, its
  spectrum is $\boldsymbol\lambda^*$.
  Inverting the depolarization gives precisely
  \eqref{eq:qutrit-min-pt-equality}, with $t=1/(1-p)$.
  Conversely, every state in that family with $t>1$ depolarizes to an
  absolutely separable maximizer at $p=1-1/t$, so both quantities equal
  the lower bound.
  The nonnegativity ranges in Corollary~\ref{cor:qutrit-min-pt}
  give the stated sharpness and strictness assertions.

  Finally, the segment from $I_d/d$ to an absolutely separable maximizer
  consists of absolutely separable states and covers every purity in
  $[1/d,M]$.  Hence both fixed-purity minima are zero throughout that
  interval.
\end{proof}

At the endpoint $P=1/n$ for $n\ge9$, the minimizing states are
$\rho=\Pi/n$, where $\Pi$ is any rank-$n$ orthogonal projection.
Here $t=4$, so
\[
  p_{\rm AS}(\rho)=p_{\rm APPT}(\rho)=\frac34.
\]

\smallskip \noindent\textbf{Two-qutrit specialization.}
For $n=3$, the matrices $K,\widetilde K$ coincide with the matrices
$L_1,L_2$ used in \cite{CK}, so their quantity $\mu_{\min}$ equals
$2q(\rho)$ by \eqref{eq:qutrit-min-pt-matrices}.
Since $M=17/121$, \eqref{eq:qutrit-min-pt-bound} becomes
\[
  2q(\rho)\le
  \frac{8-11\sqrt{2(9P-1)}}{36},
\]
which is the inequality stated in \cite[Lemma~II.5]{CK}.  Moreover, when
$\mu_{\min}<0$, \eqref{eq:qutrit-depolarizing-strength} becomes
\[
  p_{\rm APPT}(\rho)
  =\frac{9\mu_{\min}}{9\mu_{\min}-2},
\]
recovering the depolarizing-boundary formula \cite[Eq.~(17)]{CK}.  Thus
Corollary~\ref{cor:qutrit-min-pt} and
\eqref{eq:qutrit-depolarizing-purity-bound} recover the corresponding
two-qutrit statements and extend them to every qutrit--qudit system, with
all equality cases specified.

\section{Conclusion}
\label{sec:conclusion}

We have determined the exact maximum purity of APPT states on
$\BC^3\otimes\BC^n$ for every $n\ge3$, proved that the same maximum and
maximizing spectra hold for absolute separability, and thereby confirmed the
inscribed-polytope purity conjecture of Ahiable, Kothakonda, and Winter in all
qutrit--qudit dimensions.  The proof also yields a purity-dependent bound on
the smallest partial-transpose eigenvalue attainable over the global unitary
orbit, with complete equality cases.
The same equality families determine the least amount of global depolarizing
noise at fixed purity for both absolute separability and APPT throughout
the sharpness ranges.

At $n=9$, the multiplicity of the larger eigenvalue in the unique maximizing
spectrum changes from one to $n$.
For $n\ge9$, the maximum purity is $9/8$ times the maximally mixed purity
$1/d$, and the Hilbert--Schmidt circumradius is $1/\sqrt{8d}$.

For $n\ge9$, the bounds for $q(\rho)$ and the minimum amounts of
depolarizing noise are strict when $P>1/n$.  Determining their optimal
values at fixed purity throughout $1/n<P<1$ is a natural continuation of
the quantitative results.

A remaining problem is the corresponding optimization for local dimension
$m\ge4$.  In particular, it remains to determine when the inscribed-polytope
maximizers of \cite{AKW} are the true APPT purity maximizers, and whether the
resulting APPT maximum is also attained by absolutely separable states.  The
qutrit case shows that these questions can require Hildebrand inequalities
beyond the simplest outer-polytope constraint, but also that a small number of
such inequalities can suffice to determine the exact purity maximum and all
maximizing spectra.

\section*{Declarations}

\smallskip \noindent\textbf{Funding.}
The author received no funding for this work.

\medskip
\smallskip \noindent\textbf{Competing interests.}
The author has no relevant financial or non-financial interests to
disclose.

\medskip
\begin{samepage}
\smallskip \noindent\textbf{Use of AI tools.}
ChatGPT 5.6 Sol (OpenAI) was used to perform symbolic computations and
simplify some mathematical arguments, and ChatGPT 6 was used to improve
the exposition.  Computations were also
cross-checked with Mathematica.
The author verified all mathematical content and takes full responsibility
for the manuscript.
\end{samepage}

\medskip
\smallskip \noindent\textbf{Data availability.}
No datasets were generated or analyzed in this study.  The proofs
and explicit calculations supporting the results are included in
the paper.


\begin{thebibliography}{99}

  \bibitem{AV}
  J. Abellanet-Vidal, G. M\"uller-Rigat, G. Rajchel-Mieldzio\'c, and A. Sanpera,
  \textit{Sufficient criteria for absolute separability in arbitrary
  dimensions via linear map inverses}, Rep. Prog. Phys. \textbf{88} (2025),
  107601.

  \bibitem{AKW}
  J. Ahiable, N. B. T. Kothakonda, and A. Winter,
  \textit{The geometry of absolute separability and other convex matrix
  properties from spectrum}, arXiv:2608.03390v2 [quant-ph] (2026).

  \bibitem{CK}
  M. Chen and N. Klco,
  \textit{Susceptibility of qudit--qudit entanglement to quantum noise: insights
  from the negativity rank}, arXiv:2610.10820v1 [quant-ph] (2026).

  \bibitem{DK}
  H. P. D\~ung and V. T. Kh\^oi,
  \textit{On the maximum purity of absolutely separable bipartite states},
  arXiv:2510.19508v1 [math-ph] (2025).

  \bibitem{Hi}
  R. Hildebrand,
  \textit{Positive partial transpose from spectra}, Phys. Rev. A \textbf{76}
  (2007), 052325.

  \bibitem{HHH}
  M. Horodecki, P. Horodecki, and R. Horodecki,
  \textit{Separability of mixed states: necessary and sufficient conditions},
  Phys. Lett. A \textbf{223} (1996), 1--8.

  \bibitem{Jo}
  N. Johnston,
  \textit{Separability from spectrum for qubit--qudit states}, Phys. Rev. A
  \textbf{88} (2013), 062330.

  \bibitem{Kondra}
  T. V. Kondra, P. Barrios Hita, J. Neumann, H. Kampermann, and D. Bru\ss,
  \textit{Fundamental limitations on entanglement extraction from purity},
  arXiv:2605.29197 [quant-ph] (2026).

  \bibitem{Pe}
  A. Peres,
  \textit{Separability criterion for density matrices}, Phys. Rev. Lett.
  \textbf{77} (1996), 1413--1415.

  \bibitem{SC}
  Z. Song and L. Chen,
  \textit{Extreme points of sets of absolutely separable and positive partial
  transpose states}, Phys. Rev. A \textbf{112} (2025), 022409.

  \bibitem{TrS}
  A. T. Tran,
  \textit{Spectral optimization for absolutely PPT states: purity, entropy, and
  volume decay}, arXiv:2609.18568 [quant-ph] (2026).

  \bibitem{WCS}
  N. Wang, L. Chen, and Z. Song,
  \textit{Extreme points of absolutely PPT states with exactly three distinct
  eigenvalues}, Phys. Scr. \textbf{101} (2026), 315210.

\end{thebibliography}
\end{document}